# Enhanced dimerization of TiOCl under pressure: spin-Peierls - to - Peierls transition.

S. Blanco-Canosa,[1] F. Rivadulla,[1]* A. Piñeiro,[2,3] V. Pardo,[2,3] D. Baldomir,[2,3] D. I. Khomskii,[4] M. M. Abd-Elmeguid,[4] M. A. López-Quintela,[1] J. Rivas[2]

*Departamentos de [1]Química-Física y [2]Física Aplicada, Universidad de Santiago de Compostela, 15782 Santiago de Compostela, Spain.*
*[3]Instituto de Investigaciones Tecnológicas, Universidad de Santiago de Compostela, 15782 Santiago de Compostela, Spain.*
*[4]II Physikalisches Institut, Universität zu Köln, Zülpicher Str. 77, D-50937 Köln, Germany.*

*e-mail:qffran@usc.es

We report high-pressure x-ray diffraction and magnetization measurements combined with *ab*-initio calculations to demonstrate that the high-pressure optical and transport transitions recently reported in TiOCl, correspond in fact to an enhanced $Ti^{3+}$-$Ti^{3+}$ dimerization existing already at room temperature. Our results confirm the formation of a metal-metal bond between $Ti^{3+}$ ions along the b-axis of TiOCl, accompanied by a strong reduction of the electronic gap. The evolution of the dimerization with pressure suggests a crossover from the spin-Peierls to a conventional Peierls situation at high pressures.

1D electronic systems are particularly susceptible to different instabilities and to the effect of thermal/quantum fluctuations. In many cases this results in the low-lying gapless excitations that frustrate a long-range order and leads to a very rich phase diagram, with interesting phases like spin liquids or resonance valence bonded states.[1,2] Interaction among different degrees of freedom provides a way for opening up a gap for the collective excitations, and different types of long-range order become then possible. Among the latter, a magnetoelastic coupling in S=1/2 chains results in a spin-Peierls phase transition to a low temperature nonmagnetic dimerized structure,[3,4] which is a localized counterpart of a conventional Peierls transition for itinerant electrons.[5] This effect was found in $CuGeO_3$,[6] for a long time the only inorganic system with a pure spin-Peierls distortion. Recently, the possibility that TiOCl shows a low temperature spin-Peierls dimerization has been suggested, although the behavior of this material is probably more complex. Temperature dependent x-ray diffraction,[7] susceptibility,[8] ESR[9] and NMR[10] show that at $T_{SP}$= 66 K, TiOCl exhibits a first order transition with a doubling of the cell along the Ti chains (*b*-axis) to a low temperature monoclinic $P2_1/m$ phase. Below this temperature a spin singlet dimerization of the lattice opens up a magnetic gap, supporting a spin-Peierls scenario. However, the ratio of the spin-gap to the transition temperature is between three to five times larger than expected on the basis of a BCS-like mean-field model.[11] Moreover, there is a second-order phase transition at $T_{ISP}$=91 K corresponding to an incommensurate dimerized state due to frustrated interchain interactions.[8]

On the other hand, Kuntscher *et al.*[12] observed a strong suppression of the transmittance and an abrupt increase of the near-infrared reflectance above ~10 GPa in TiOCl. These effects were interpreted as a pressure-induced metallization of the 1D chain. This finding, if confirmed, could provide an exceptional opportunity to study the low energy electronic excitations of a 1D itinerant-electron system, as well as a type of Mott transition in reduced dimensions. Later on, a structural phase transition was observed in isostructural TiOBr, coincident with the proposed pressure-induced metallization.[13] Nevertheless, the structure of this high-pressure phase was not resolved, and the issue of whether the system is truly metallic at high pressure was not clarified. Recently, Forthaus *et al.*[14] measured the resistivity up to ~25 GPa and reported an anomalous decrease of the electronic gap above ~12 GPa in TiOCl, but ruled out the existence of a metallic phase at high-pressure. The reduction of the gap was attributed to a pressure-dependent competition between direct and indirect hopping along Ti-Ti and Ti-O-Ti bonds, respectively, and has been shown to be connected with an anomalous change of the lattice parameters with pressure. However, no evidence for structural phase transition in TiOCl has been reported within the accuracy of the energy dispersive x-ray data.

In this work, we report the strong enhancement of a dimerization of the 1D spin-chain of TiOCl under pressure, which for P>~10 GPa exists already at room temperature. Ab-initio calculations confirm the dramatic decrease of the charge gap in the high-pressure dimerized phase, from ~ 1.5 eV down to ~ 0.3 eV, in very good agreement with both transport and optical experiments. Our results demonstrate the formation of direct $Ti^{3+}$-$Ti^{3+}$ bonds at room temperature and high pressure, with the shortest distance being consistent with partial electronic delocalization along this bond. Thus we can interpret a rapid change of properties of TiOCl at around 10 GPa as a crossover from the spin-Peierls to an ordinary Peierls dimerization.

Single crystal samples of TiOCl were grown by chemical vapor transport at 650-700 ºC in evacuated quartz tubes from stoichiometric amounts of Ti, $TiO_2$, $TiCl_3$.[15] Magnetic susceptibility under pressure up to 10 kbar was measured in a SQUID magnetometer using a commercial Be-Cu cell from EasyLab. X-ray diffraction under pressure (up to 15 GPa) were done at Station 9.5HPT (λ=0.44397 Å) of the Daresbury Synchrotron Radiation Source,[16] using diamond anvil cells and a 4:1 $CH_3OH$:$C_2H_5OH$ mixture as pressure media. Ab-initio calculations were performed based on the density functional theory, with the WIEN2k software[17] using a full-potential, all-electron scheme based on the APW+lo method.[18] Strong correlations were introduced by means of the LDA+U method.[19]

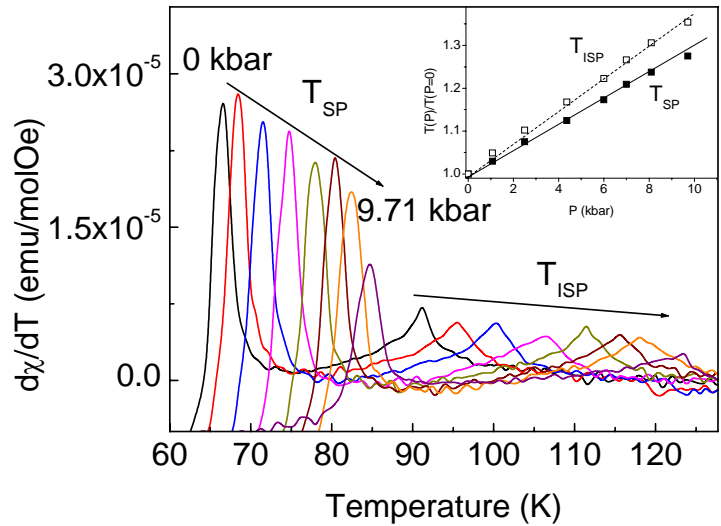

Fig. 1: Derivative of the temperature dependence of the magnetic susceptibility curves at different pressures. The precise evolution of $T_{ISP}$ and $T_{SP}$ is shown in the inset.

In Fig. 1 we show the derivative of the magnetic susceptibility vs. temperature and its pressure dependence up to ~10 kbar. Two distinct phase transitions, at the incommensurate spin-Peierls temperature ($T_{ISP}$ ~ 91(1) K) and at the commensurate spin-Peierls temperature ($T_{SP}$ ~ 66(1) K), are observed. The pressure dependence of the magnetic transition temperature derived from these measurements are $(\partial \ln T_{SP}/\partial P)=2.88\times10^{-1}$ $GPa^{-1}$ and $(\partial \ln T_{ISP}/\partial P)=3.64\times10^{-1}$ $GPa^{-1}$. These values are slightly smaller to those reported for TiOBr $(\partial \ln T_{SP}/\partial P)=3.4\times10^{-1}$ GPa$^{-1}$,[20] and to our own measurements $(\partial \ln T_{SP}/\partial P)=3.5\times10^{-1}$ GPa$^{-1}$ and $(\partial \ln T_{ISP}/\partial P)=4.7\times10^{-1}$ GPa$^{-1}$.

In Fig. 2 we show the results of the x-ray diffraction under pressure. Increasing pressure up to ~9.5 GPa causes a progressive displacement of the peaks towards higher angles and a broadening, most likely due to a slight non-hydrostaticity at the higher pressures. The pattern at ambient pressure and up to 9.5 GPa are well fit within the orthorhombic symmetry of the *Pmmn* space group, giving a=3.79(4) Å, b=3.38(3) Å and c=8.03(7) Å, at room temperature. Fitting the pressure dependence of the volume below 9.5 GPa to the Birch-Murnaghan equation gives a bulk modulus B= 67(4) GPa. The orthorhombic crystal structure of TiOCl consists of 2D Ti-O bilayers along the *ab* plane well separated by Cl ions along the *c* direction.[21] Consistent with this structure, the variation of the lattice parameters is strongly anisotropic and the bulk modulus is dominated by the highly compressible *c*-axis. For the *ab* plane, where the relevant exchange interactions occur, we found that the *b*-axis is much more compressible ($B_b$=130±10 GPa) than the *a*-axis ($B_a$=590±30 GPa), in agreement with previous reports.[14]

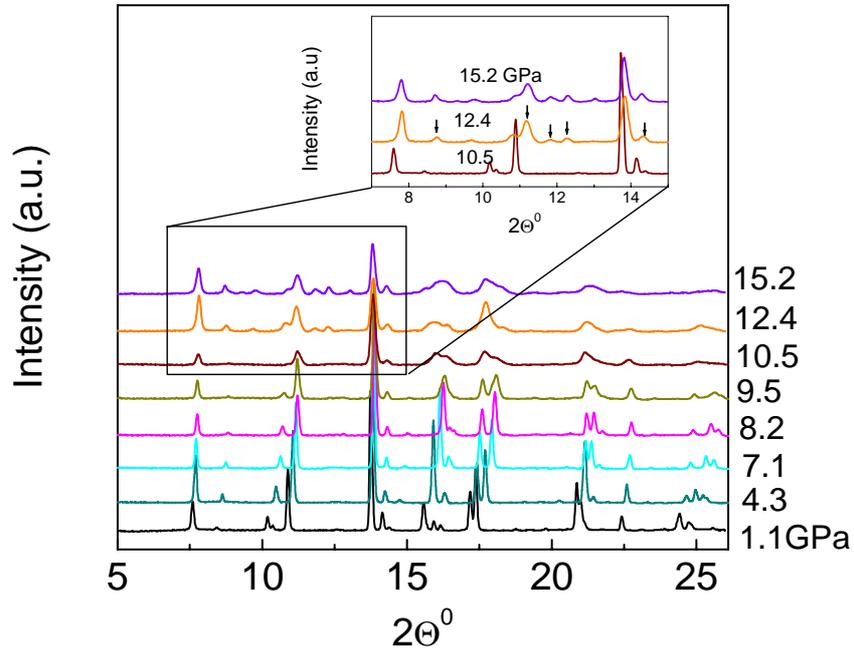

Fig. 2: Evolution of the x-ray diffraction pattern under pressure. The inset shows the detail of the high-pressure transformation, with the new peaks of the monoclinic *P2$_1$/m* marked by arrows.

At pressures ~ 10 GPa, where the optical[12] and transport[14] data have shown a strong change of behavior, with the rapid decrease of an energy gap, our high-resolution x-ray diffraction using synchrotron radiation show a structural phase transition (inset in Fig. 2). The peaks of the high-pressure phase can be indexed within the monoclinic space group $P2_1/m$. Rietveld refinement of the high pressure monoclinic phase (see Fig. 3) results in values of the lattice parameters at 15.2 GPa of $a$= 3.536(3) Å, $b$= 6.631(3) Å, $c$= 7.027(7) Å, and $\beta$ = 98.89(8)°. The small differences between the experimental and the calculated patterns are due to the difficulty in reproducing the broadening of the peaks. However, the structural transition is reproducible and reversible, ruling out any extrinsic origin. These results are also consistent with those reported for TiOBr.[13]

According to our structural data, the effect of pressure is to induce an alternating tilting of the $TiO_4Cl_2$ octahedra resulting in a doubling of the unit cell along the $b$ direction in the high-pressure monoclinic phase with respect to the low pressure orthorhombic structure. This results in two inequivalent $Ti^{3+}$ sites along the $b$-axis with alternating $Ti^{3+}$-$Ti^{3+}$ distances of 2.85(5) Å and 3.55(5) Å (see the inset to Fig. 3). The shortest distance is comparable to the close contact distance in Ti metal at room temperature (2.896 Å) and to the shortest distance in the dimerized phase of the cubic spinel $MgTi_2O_4$ (2.853(7) Å).[22] This result supports the formation of a metal-metal bond with partial electronic delocalization between $Ti^{3+}$ ions along the $b$-axis in TiOCl. Moreover, while in the low temperature dimerized phase at ambient pressure the difference between the short-long bonds is roughly 5%, our x-ray data shows that this difference increases dramatically up to ≈20% in the room-temperature high-pressure dimerized phase.

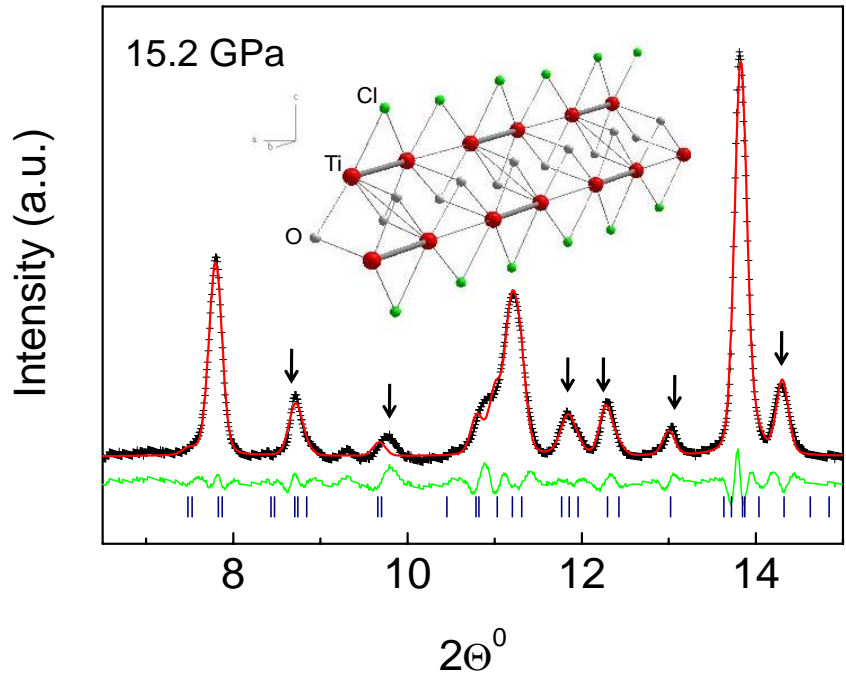

Fig. 3: Rietveld refinement of the high-pressure monoclinic phase (+ are the experimental points and solid red line is the fitting). Vertical arrows mark the new peaks of the high-pressure monoclinic phase. The inset shows the structure generated from the refinement, highlighting the short Ti-Ti bonds along the *b*-axis, which are displaced half a period between the chains.

The evolution of the dimerized ground state for the monoclinic phase with pressure was reproduced by *ab*-initio calculations. They were performed at several pressures utilizing the lattice parameters measured experimentally and, for each pressure, optimizing the internal atomic parameters. For this structural optimization we used the GGA scheme,[23] minimizing the total energy of the system and reducing the forces on the atoms below 2 mRy/a.u. For the lattice parameters above 10 GPa, the structure optimization within a monoclinic structure leads to the formation of $Ti^{3+}$-$Ti^{3+}$ dimers along the *b* axis, confirming the experimental results. At P= 15 GPa, the calculated alternating $Ti^{3+}$-$Ti^{3+}$ distances are 2.95±3 Å and 3.69±4 Å. Moreover, the structural transition produces changes in the electronic structure with the formation of molecular orbitals within the dimer. This modifies strongly the excitation spectra, leading to a dramatic reduction of the electronic gap with respect to the undimerized orthorhombic phase (Figure 4). Within the orthorhombic structure the gap decreases with pressure by about 30 %, to ~ 1 eV, in good agreement with the reduction observed exerimentally.[12,14] In the dimerized monoclinic structure, stable at room temperature beyond ~10 GPa, the gap reduces drastically to ~0.3 eV. This is again in very good agreement with the experimental observation of a sudden decrease of the charge gap at room temperture above ~10

GPa.[12] Our calculations predict that more than 30 GPa would be necessary to obtain such a reduction of the electronic gap without considering a structural transition.

On the other hand, in both phases the electronic structure of TiOCl derived from our calculations is strongly anisotropic (quasi-one-dimensional). The occupied $t_{2g}$ orbital ($d_{yz}$) of the $Ti^{3+}$ ($d^1$) ions shows a large hopping integral along the $b$ direction of the crystal. We have calculated total energies within the LDA+U approximation for various magnetic configurations and several pressures within the orthorhombic structure, and fit them to a Heisenberg Hamiltonian. Due to the structure of the material, we considered three different magnetic couplings and their pressure dependence: $J_b$, within the 1D chains along the $b$ axis; $J_s$, between a Ti and its 4 closest neighbors; and $J_a$, the coupling along the $a$ axis. At ambient pressure, the ratio $|J_b/J_a|\sim300$ and $|J_b/J_s|\sim25$, and therefore the direct AF coupling along the $b$-axis is the dominant exchange interaction in TiOCl. This conclusion is confirmed by the lower transition temperatures in TiOBr, with an almost identical $a$ parameter but with a ~5% reduction of $b$ with respect to TiOCl.[24]

On the other hand, the variation of $J_b$ with pressue from our calculations is ~1.4 times faster that $J_a$, and even faster with respect to the other directions. This corroborates that the pressure dependence of the transport in this system is dominated by the direct $Ti^{3+}$-$Ti^{3+}$ hopping, and more important, the 1D character of the compound increases with pressure. Fitting the results to a quasi-1D-Heisenberg model yields a value for the transition temperature at ambient pressure of ~50 K and a pressure dependence of $(\partial \ln T/\partial P)=4.6\times10^{-1}$ $GPa^{-1}$, close to the experimental values.

The pressure dependence measured for $T_{SP}$ and $T_{ISP}$ is much larger than expected on the basis of superexchange theory for a conventional antiferromagnetic transition. Partial electronic delocalization along the Ti-Ti bonds results in an effective screening of the Coulomb interaction and a decrease of U, which then increases the rate of variation of the exchange interaction with pressure as observed in other systems close to the itinerant boundary.[25]

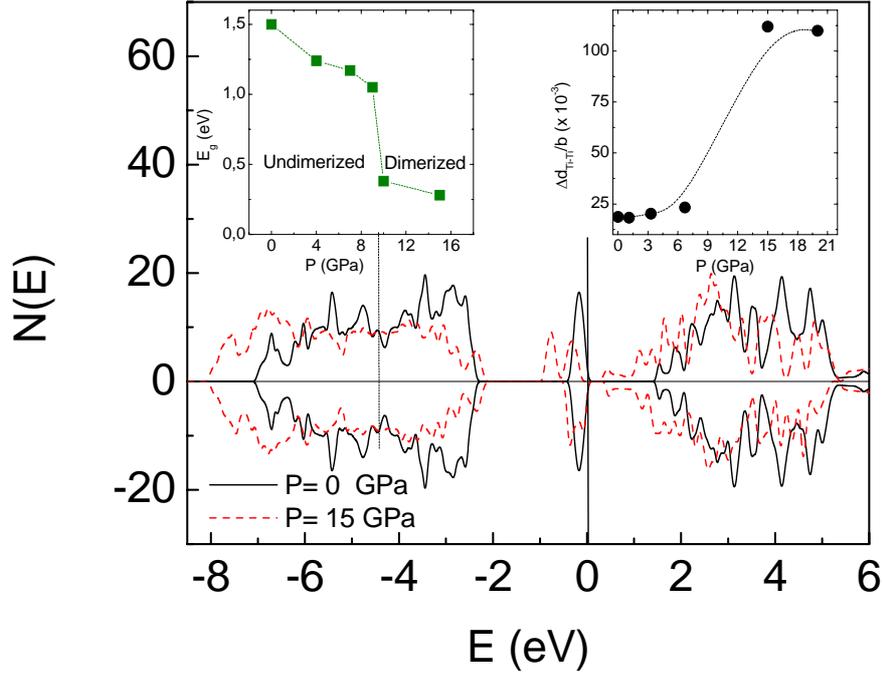

Fig. 4: Density of states for TiOCl at ambient pressure (*Pmmn*) and at P= 15 GPa (*P2$_1$/m*). The vertical line marks the Fermi energy. The insets show the evolution of the band gap (left) and the difference between the short and long distances in the dimers (right) with pressure.

An important consequence follows also from the non-monotonic evolution of the degree of dimerization (difference of short-long distances) in the dimerized chain with pressure. As observed in the inset to Fig. 4, this difference increases dramatically between 10 and 15 GPa. In this pressure range, the short Ti-Ti distance approaches the critical distance for metallic bonding. Our results support an enhanced shortening of the bond in order to form a molecular orbital, i.e. by a purely electronic mechanism. In this case the magnetic structure will play a secondary role, and hence the high pressure phase departs from the spin-Peierls scenario, and may be rather interpreted as a usual Peierls dimerization, proceeding from the itinerant limit. Since in TiOBr not only the change of the optical and structural properties is very similar to that of TiOCl,[13] but also $\partial \ln T_{SP}/\partial P$ is strongly enhanced with pressure,[20] we think that our scenario of a pressure-induced crossover from the spin-Peierls to a conventional Peierls state should be also valid for TiOBr.

In summary, we have demonstrated the existence of a structural transition at high pressure in TiOCl, at which the $Ti^{3+}$-$Ti^{3+}$ dimerization is strongly enhanced as compared with that at ambient pressure. The shortest $Ti^{3+}$-$Ti^{3+}$ metal-metal distance is consistent with partial electronic delocalization along the pairs. A dramatic reconstruction of the electronic crystal structure is driven by the proximity to the itinerant electron boundary in the short Ti-Ti bonds, suggesting a completely different mechanism for the high-pressure dimerization, which should be treated not as a spin-Peierls, but rather as a Peierls transition.

**Acknowledgements.**
Dr. Alistair Lenie is acknowdeged for his help during the measurements at the Daresbury Synchrotron Radiation Facility. Financial support from MEC (MAT2006-10027, MAT2007-66696-C02-02 and HA2006-0119) of Spain and Xunta de Galicia (PXIB20919PR) is acknowledged. F. R. and S. B-C. also thank MEC of Spain for support under Ramón y Cajal and FPU programs respectively. M.M.A and D.I.K. would like to thank the Deutsche Forschungsgemeinschaft (DFG) via SFB 608 for financial support.